\documentclass[twocolumn]{aastex63}
\usepackage{times}
\usepackage{amsmath}
\usepackage{textcomp}
\usepackage{amssymb}
\usepackage{natbib}
\usepackage{float}
\usepackage{color}
\usepackage{graphicx}

\newcommand{\msun}{\ensuremath{M_{\odot}}}
\newcommand{\lum}{erg\,s$^{-1}$}
\newcommand{\fermi}{{\it Fermi}}

\newcommand{\ergflux}{\mbox{${\rm \, erg \,\, cm^{-2} \, s^{-1}}$}}
\newcommand{\gm}{$\gamma$}

\newcommand{\Ca}{Ca{\sevenrm II}}

\newcommand{\OIIIb}{[O{\sevenrm\,III}]\,$\lambda$5007}

 \font\sevenrm=cmr7 scaled 1000

\received{\today}
\revised{\today}
\accepted{\today}
\submitjournal{ApJ}

\shorttitle{Gamma-ray Emission of FR0 Galaxies}
\shortauthors{Pannikkote, Paliya, and Saikia}

\begin{document}
\title{Hunting Gamma-ray emitting FR0 radio galaxies in wide-field sky surveys}

\correspondingauthor{Vaidehi S. Paliya}
\email{vaidehi.s.paliya@gmail.com}

\author[0009-0004-8724-9163]{Meghana Pannikkote}
\affiliation{Department of Physics, National Institute of Technology Calicut, Calicut-673601, India}
\author[0000-0001-7774-5308]{Vaidehi S. Paliya}
\affiliation{Inter-University Centre for Astronomy and Astrophysics (IUCAA), SPPU Campus, Pune 411007, India}
\author[0000-0002-4464-8023]{D. J. Saikia}
\affiliation{Inter-University Centre for Astronomy and Astrophysics (IUCAA), SPPU Campus, Pune 411007, India}

\begin{abstract}
The latest entry in the jetted active galactic nuclei (AGN) family is the Fanaroff-Riley type 0 (FR0) radio galaxies. They share several observational characteristics, e.g., nuclear emission and host galaxy morphology, with FR I sources; however, they lack extended, kiloparsec-scale radio structures, which are the defining features of canonical FR I and II sources. Here we report the identification of 7 \gm-ray emitting AGN as FR0 radio sources by utilizing the high-quality observations delivered by ongoing multi-wavelength wide-field sky surveys, e.g., Very Large Array Sky Survey. The broadband observational properties of these objects are found to be similar to their \gm-ray undetected counterparts. In the \gm-ray band, FR0 radio galaxies exhibit spectral features similar to more common FR I and II radio galaxies, indicating a common \gm-ray production mechanism and the presence of misaligned jets. Although the parsec-scale radio structure of FR0s generally exhibits a wide range, with about half having emission on opposite sides of the core, the \gm-ray detected FR0s tend to have dominant cores with core-jet structures. We conclude that dedicated, high-resolution observations are needed to unravel the origin of relativistic jets in this enigmatic class of faint yet numerous population of compact radio sources.
 
\end{abstract}


\section{Introduction}
The viewing angle of the relativistic jet with respect to the observer's line of sight plays a decisive role in explaining the physics of jetted active galactic nuclei (AGN). For smaller viewing angles ($\theta_{\rm v}\lesssim 1/\Gamma$, $\Gamma$ is the bulk Lorentz factor), the radiation emitted from the jet is strongly amplified due to relativistic beaming effects \citep[cf.][]{1966Natur.211..468R} and such AGN are called blazars. Blazars often exhibit compact core-jet morphology, superluminal motion, and high-brightness temperature at radio frequencies. At large viewing angles, on the other hand, spectacular $\sim$kpc-Mpc scale relativistic jets have been observed at radio wavelengths leading to the identification of such AGN as radio galaxies. Historically, these objects have been classified as Fanaroff-Riley (FR) type I and II sources based on their 178 MHz luminosity and distinctive radio morphology \citep[][]{1974MNRAS.167P..31F}. However, several outliers have been reported in recent observations with the Low-Frequency Array \citep[LOFAR,][]{2019MNRAS.488.2701M}. Furthermore, the misalignment of the jet implies a relatively weak Doppler boosting, thereby making radio galaxies faint \gm-ray emitters. Indeed, they form a tiny fraction ($<$2\%) of all \gm-ray emitting AGN detected with the \fermi~Large Area Telescope \citep[LAT,][]{2022ApJS..263...24A}.

With the advent of sensitive, high-resolution radio surveys, e.g., Faint Images of the Radio Sky at Twenty centimeters survey \citep[FIRST,][]{2015ApJ...801...26H}, it is becoming evident that a majority of known radio-loud AGN in the nearby Universe ($z\lesssim$0.1) are compact with linear sizes $\lesssim$10 kpc \citep[][]{2009A&A...508..603B,2012MNRAS.421.1569B}. Such objects are termed as FR0 radio galaxies \citep[][]{2011AIPC.1381..180G}. Interestingly, their optical spectroscopic properties and host galaxies, which are red, early-type, low-excitation galaxies (LEGs), are similar to FR Is, and the only distinctive feature is the paucity of extended, i.e., $\sim$kpc-scale, radio jets commonly observed from FR I and FR II radio galaxies. \citet[][]{2018A&A...609A...1B} compiled a catalog of such sources, also known as {\it FR0CAT}, by utilizing observations from the FIRST and Sloan Digital Sky Surveys \citep[SDSS, see also,][]{2012MNRAS.421.1569B}. A redshift filter of $z\leq$0.05 was adopted to select AGN with linear sizes $<$5 kpc. The multi-wavelength follow-up of these sources revealed the presence of mildly relativistic flow of the plasma and similarity with FR I sources in the X-ray band as well \citep[cf.][]{2018ApJ...863..155C,2018MNRAS.476.5535T,2020A&A...642A.107C}. The properties and nature of FR0 radio galaxies have been reviewed recently by \citet{2023A&ARv..31....3B}.

A few of the FR0 radio galaxies have also been detected with the \fermi-LAT, and the stacking of the \gm-ray data of the undetected FR0 population has revealed them to be \gm-ray emitters as a whole \citep[][]{2016MNRAS.457....2G,2021ApJ...918L..39P}. This faint yet numerous population has been proposed as plausible sites of ultra-high-energy cosmic-ray acceleration and a promising candidate for neutrino emission \citep[][]{2018MNRAS.475.5529T,2021APh...12802564M}. Therefore, it is imperative to increase the sample size of \gm-ray emitting FR0 radio galaxies to understand the radiative process powering their jets and the surrounding environment in which they are launched and grow.

\citet[][]{2021ApJ...918L..39P} analyzed the \gm-ray data of sources present in {\it FR0CAT} and reported the discovery of 3 FR0 radio galaxies in the \gm-ray band\footnote{One of the \gm-ray detected FR0 galaxies, LEDA 57137, turned out to be an FR I source \citep[][]{2019MNRAS.482.2294B} leaving only two bona fide \gm-ray emitting FR0 known as of now.}. Here we adopted the reverse approach to identify new \gm-ray emitting FR0s. In other words, we prepared a sample of known \gm-ray emitting AGN from the third data release of the fourth catalog of the \fermi-LAT detected AGN \citep[4LAC-DR3,][]{2022ApJS..263...24A} and carefully examined their optical and radio properties to identify potential FR0 radio galaxies. In this article, we report the identification of 7 \gm-ray emitting AGN as FR0 radio galaxies, thus $\sim$quadrupling the sample size of the known \gm-ray detected FR0 radio galaxies. 

Section~\ref{sec2} describes the sample selection and catalogs used to identify candidate FR0 sources. We present our findings in Section~\ref{sec3} and discuss and summarize them in Sections~\ref{sec4} and \ref{sec5}, respectively. Throughout, a flat cosmology with $H_0 = 70~{\rm km~s^{-1}~Mpc^{-1}}$ and $\Omega_{\rm M} = 0.3$ was adopted.

\section{Sample Selection and Multi-wavelength Catalog Search}\label{sec2}
We started with the 4LAC-DR3 catalog, which lists 3814 \gm-ray detected AGN, including 792 flat spectrum radio quasars, 1458 BL Lac objects, 71 non-blazar AGN, and 1493 blazar candidates of uncertain type \citep[BCU,][]{2022ApJS..263...24A}. Since the identification of FR0 sources requires an analysis of their optical spectra, we cross-matched the 4LAC-DR3 catalog with SDSS data-release 17 \citep[SDSS-DR17,][]{2022ApJS..259...35A} using a 5 arcsec search radius. To obtain the optical spectra of \gm-ray sources not lying within the SDSS footprint, we also considered the results published by \citet[][]{2021ApJS..253...46P} where they collected and analyzed the optical spectra of 1077 \fermi-LAT detected blazars. This led to the selection of 928 objects from SDSS-DR17 and 539 sources from \citet[][]{2021ApJS..253...46P}. These 1467 \gm-ray detected AGN with available optical spectra formed the parent sample to identify potential FR0 radio galaxies among them.

The main defining property of FR0 sources is the lack of extended radio emission. \citet[][]{2018A&A...609A...1B} used the data from the FIRST survey, which has an angular resolution of 5 arcsec at 1400 MHz. In order to identify genuine radio compact objects, they applied a redshift cutoff of $z\leq0.05$, which corresponds to an upper limit on the linear size of $\sim$5 kpc of the radio source considering the FIRST survey resolution. We have, on the other hand, considered the data from the Very Large Array Sky Survey (VLASS), which is in the frequency range of 2$-$4 GHz, and has a higher angular resolution ($\sim$2.5 arcsec) and a slightly higher sensitivity (0.12 mJy per beam for the first epoch of VLASS) compared to the FIRST survey \citep[][]{2020PASP..132c5001L}. To identify the VLASS-detected radio sources with linear sizes $\lesssim$5 kpc, we apply a redshift cutoff of $z\leq$0.11 on the parent sample. This exercise led to the selection of 108 radio sources. The VLASS quick-look images were downloaded\footnote{\url{http://cutouts.cirada.ca}} and visually inspected by all authors individually. This led to the rejection of 20 objects whose VLASS images exhibit an extended radio structure. We also checked the FIRST survey images of these objects, when available, for consistency. The remaining 88 sources were then subjected to optical spectroscopic analysis, as discussed below.

The FR0 sources are LEGs with their optical spectra showing a red continuum with strong absorption features arising from the host galaxy stellar population and devoid of broad emission lines \citep[cf.][]{2010A&A...519A..48B}. Therefore,  50 out of 88 sources whose optical spectra consist of a featureless blue continuum usually observed from BL Lac objects were filtered out.  The optical spectra of the remaining 38 sources were analyzed to identify potential FR0 radio galaxies among them.

\section{Results}\label{sec3}
\begin{table*}
\caption{The \gm-ray spectral parameters (adopted from the 4LAC-DR3 catalog) and other derived quantities for the \gm-ray emitting FR0 radio galaxies. The column information are as follows: (1) 4FGL name; (2) association; (3) redshift; (4) \gm-ray flux (in units of 10$^{-12}$ \ergflux); (5) \gm-ray photon index; (6) logarithmic \gm-ray luminosity (in \lum); (7) significance of the \gm-ray detection ($\sigma$); (8) logarithmic \OIIIb~line luminosity (in \lum); (9) logarithmic black hole mass (in M$_{\odot}$); (10) absolute $r$-band magnitude; and (11) Dn(4000) index.\label{tab:basic_info}
}
\begin{center}
\begin{tabular}{llccccccccc}
\hline
4FGL name  & Association & $z$ & $F_{\gamma}$ & $\Gamma_{\gamma}$ & $L_{\gamma}$ &  Det. Signif. & $L_{\rm OIII}$ & $M_{\rm BH}$ & $M_{\rm r}$  & Dn(4000)\\
(1) & (2)& (3) & (4) & (5) & (6) & (7) & (8) & (9) & (10)& (11)\\ 
\hline
J0003.2+2207	 & LEDA 1663156 & 0.099  & 1.57$\pm$0.35 & 2.14$\pm$0.16   & 43.60 & 5.5   & 39.44 & 8.01  &  $-$21.70  & 1.72 \\
J0154.3$-$0236 & LEDA 144405 & 0.082   & 2.11$\pm$0.32 & 2.17$\pm$0.12   & 43.56 & 8.7   & 40.40 & 9.41 &  $-$22.80  & 1.95 \\
J0312.4$-$3221	& NVSS J031234$-$322315 & 0.067   & 1.40$\pm$0.26 & 2.22$\pm$0.14   & 43.19 & 6.9   & 40.54 & 8.72   &  $-$21.84  & 1.72 \\
J1212.1+6412	& LEDA 2665658            &  0.108  & 2.14$\pm$0.41 & 2.57$\pm$0.13   & 43.84 & 5.4   & 40.51 & 8.43  & $-$22.50  & 1.89 \\
J1530.3+2709	& LEDA 55267              & 0.032   & 0.93$\pm$0.26 & 2.01$\pm$0.20   & 42.36 & 4.8   & 39.77 & 8.18  & $-$21.41  & 1.96 \\
J1612.2+2828	& TXS 1610+285            &  0.053  & 1.17$\pm$0.29 & 2.14$\pm$0.17   & 42.90 & 4.9   & 39.90 & 8.30  & $-$22.59  & 2.03 \\
J1612.4$-$0554	& LEDA 1038366            &  0.029  & 2.72$\pm$0.66 & 2.52$\pm$0.15   & 42.72 & 4.0   & 38.98 & 8.64   & $-$21.06  & 3.74 \\
\hline
\end{tabular}
\end{center}
\end{table*}

\begin{figure}
\vbox{
    \includegraphics[width=\linewidth]{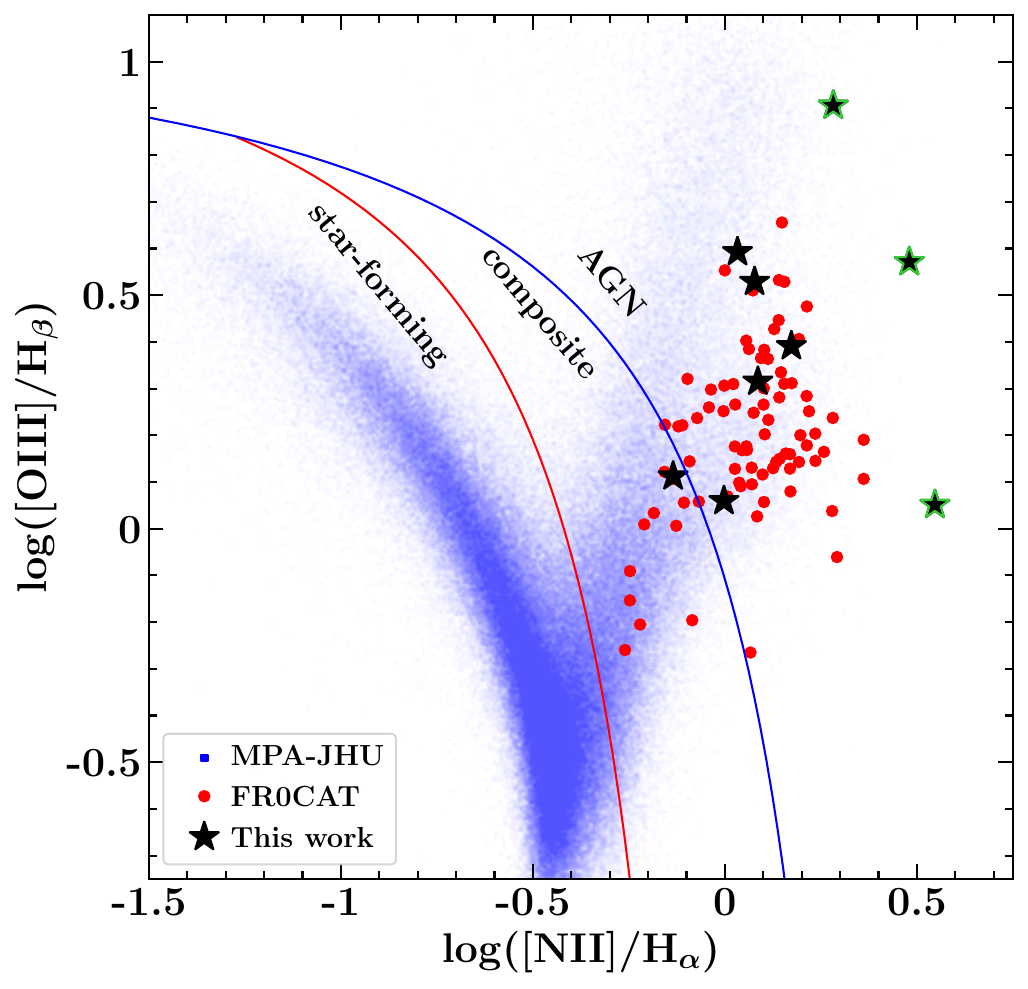}
    }
\caption{The BPT diagnostic diagram of the {\it FR0CAT} sources (red circles) and objects studied in this work (black stars). For a comparison, we also show SDSS galaxies studied by the group from the Max Planck Institute for Astrophysics, and The Johns Hopkins University \citep[MPA-JHU,][]{2004MNRAS.351.1151B}. The red and blue lines show the relations proposed to classify emission-line galaxies \citep[][]{2006MNRAS.372..961K}. Black stars with green boundaries are rejected from the analysis.} \label{fig:bpt}
\end{figure}

\begin{figure*}
\hbox{
    \includegraphics[scale=0.23]{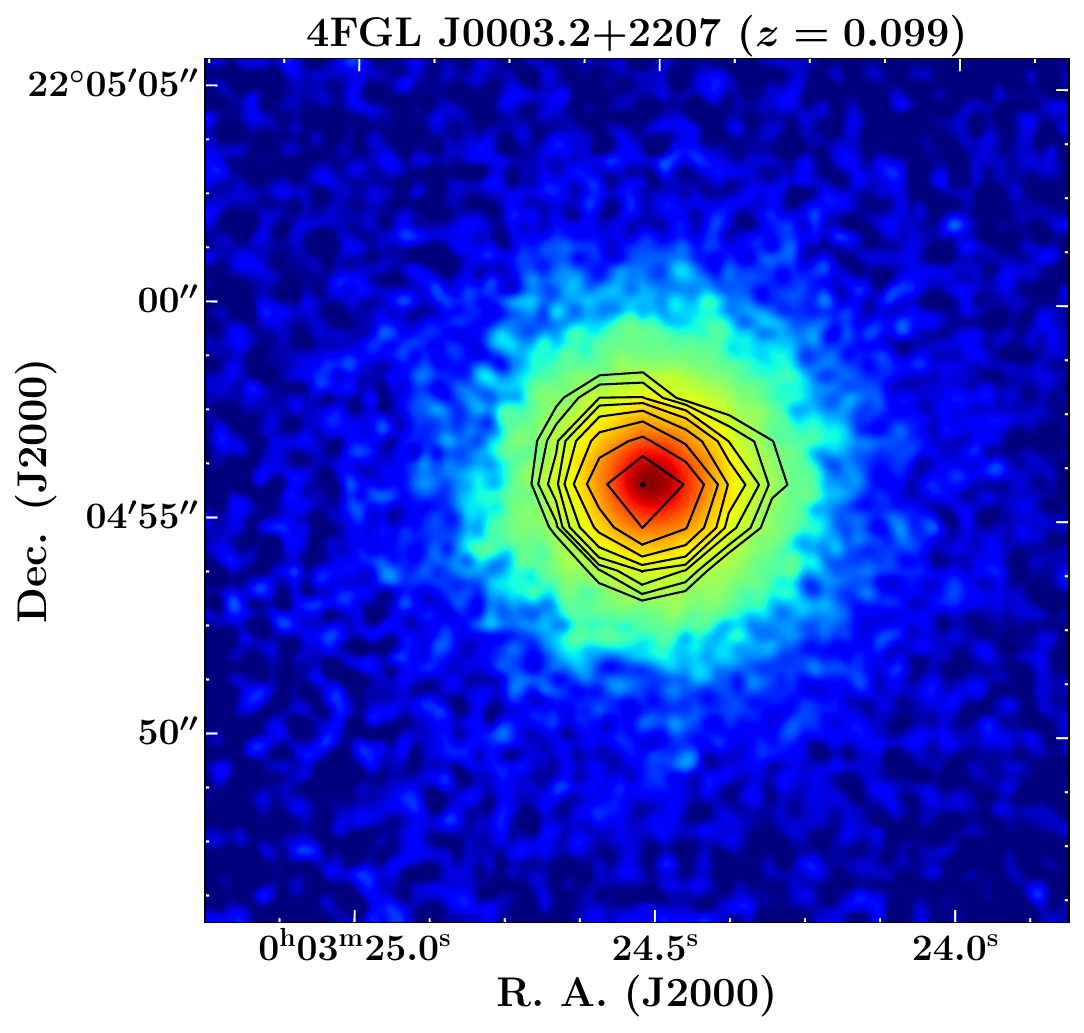}
    \includegraphics[scale=0.23]{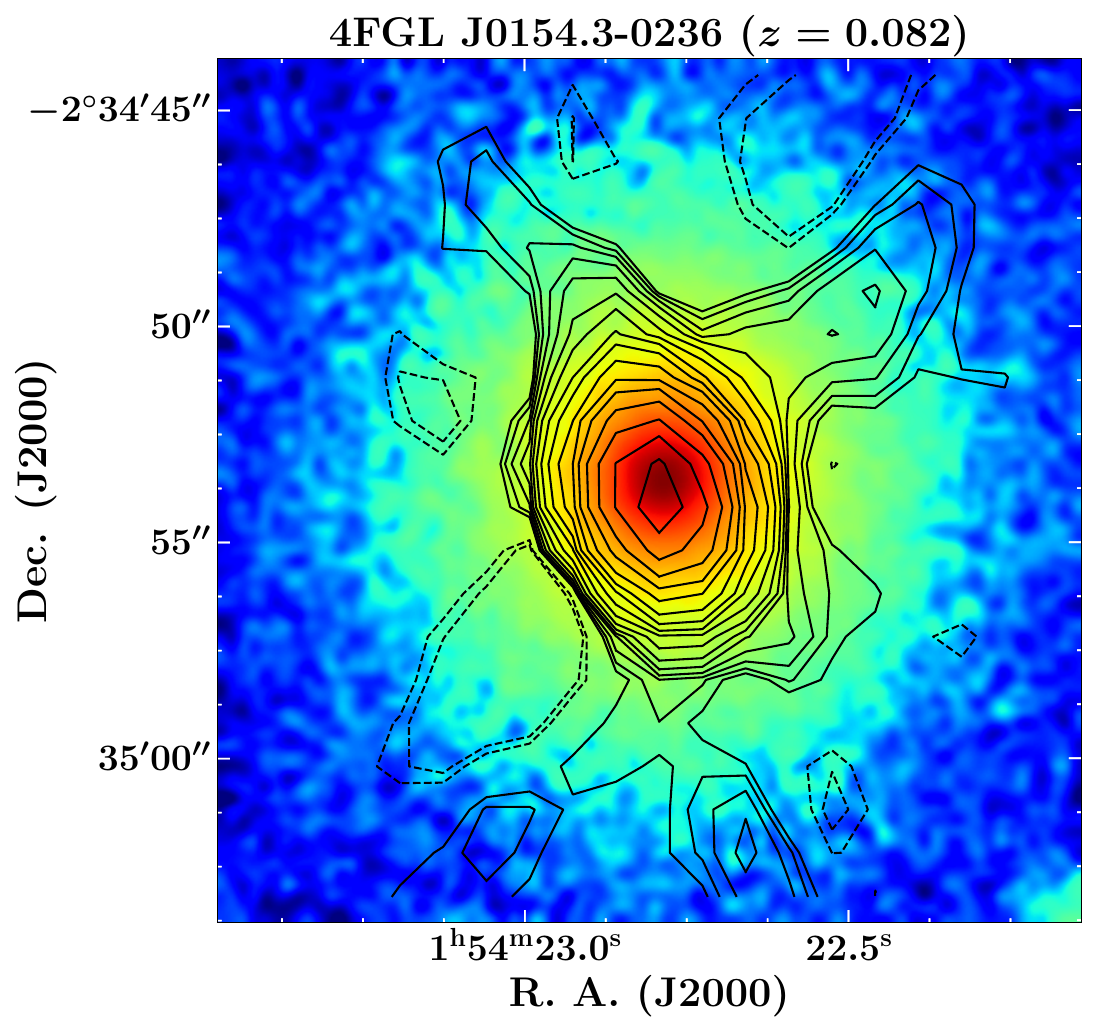}
    \includegraphics[scale=0.23]{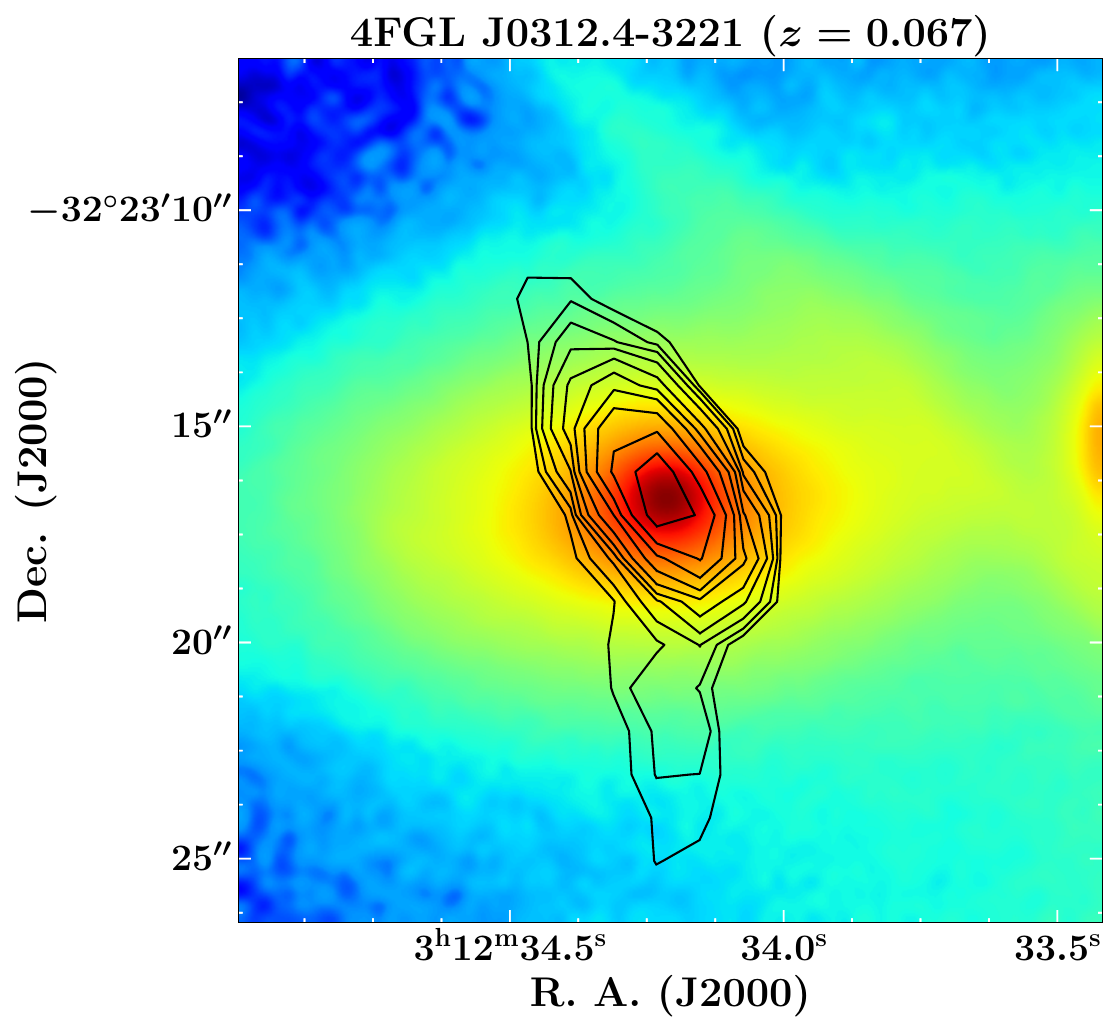}
    \includegraphics[scale=0.232]{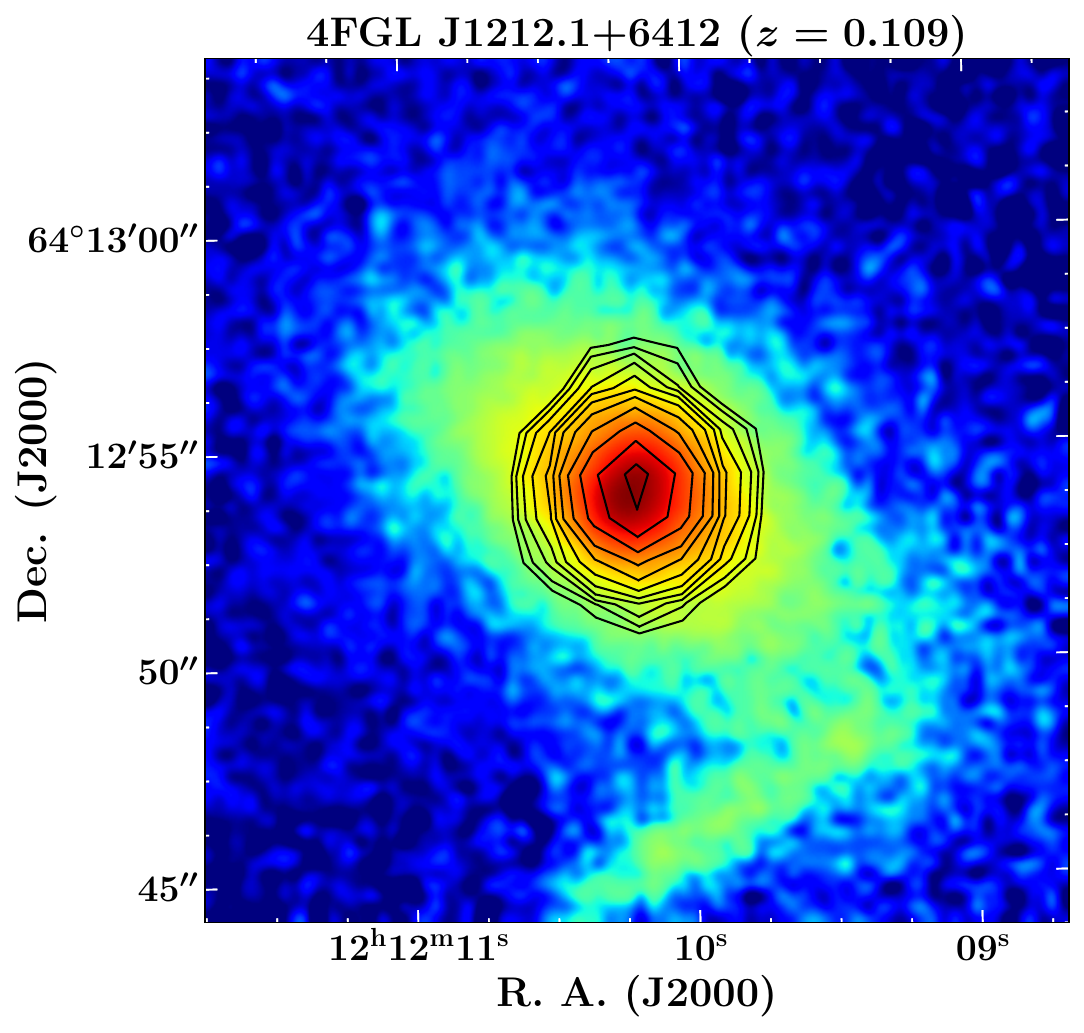}
    }
\hbox{
    \includegraphics[scale=0.25]{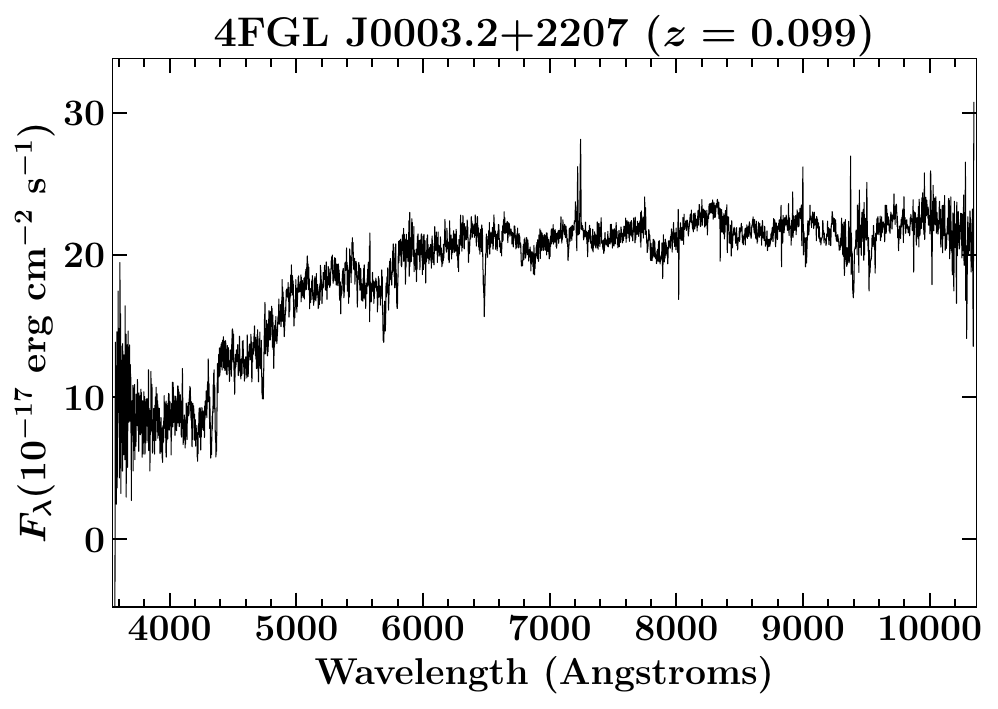}
    \includegraphics[scale=0.25]{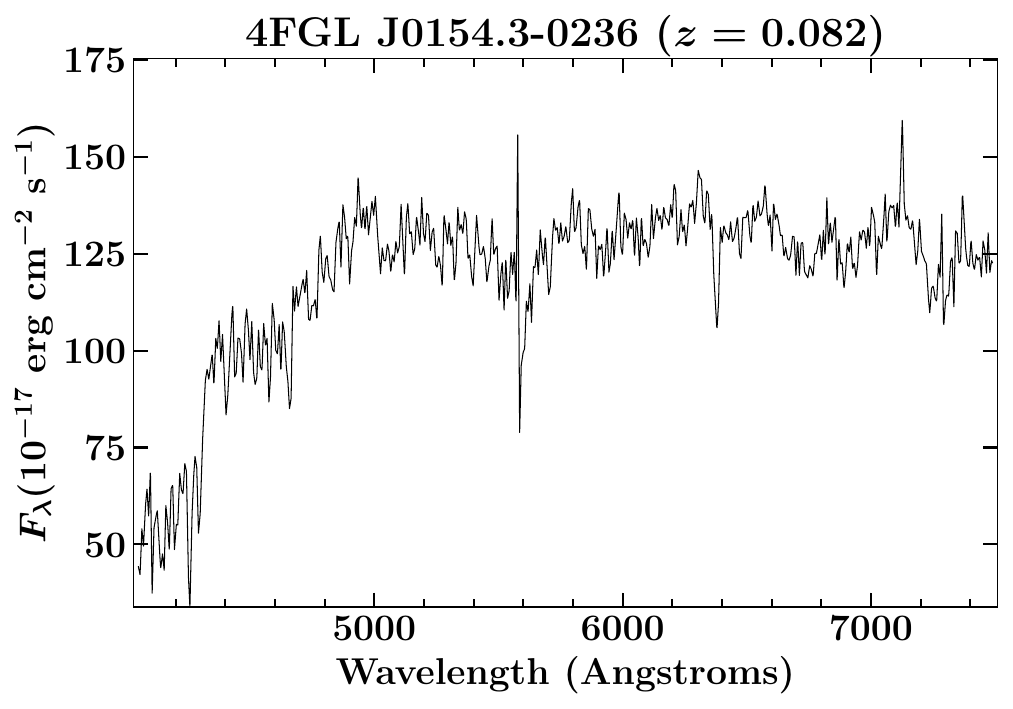}
    \includegraphics[scale=0.25]{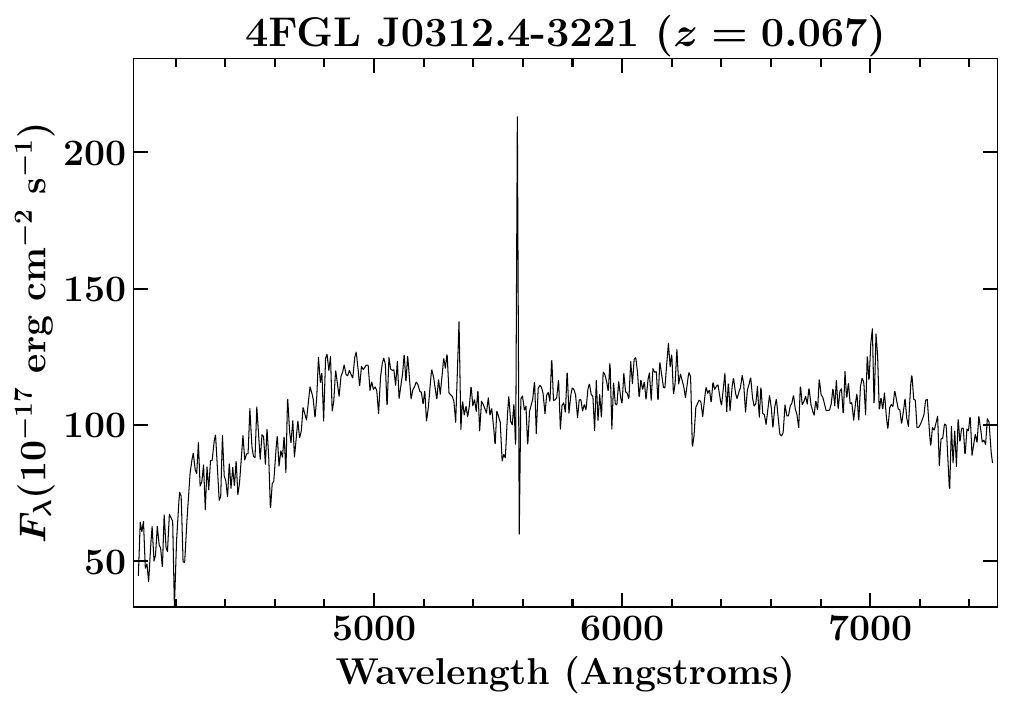}
    \includegraphics[scale=0.25]{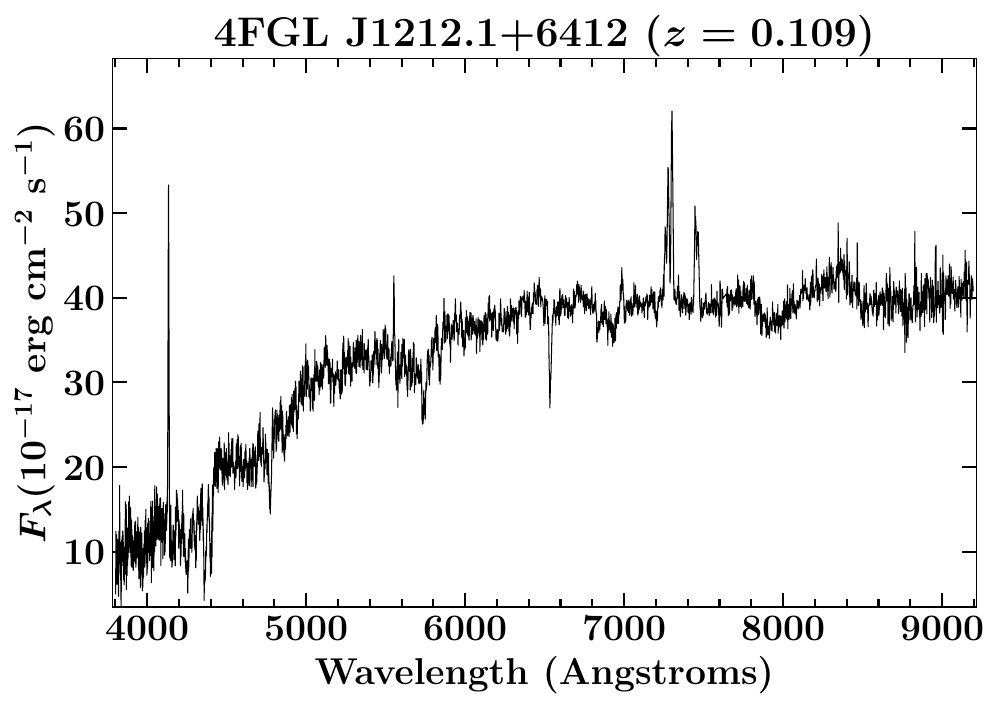}
    }
        \hbox{\hspace{2.5cm}
    \includegraphics[scale=0.23]{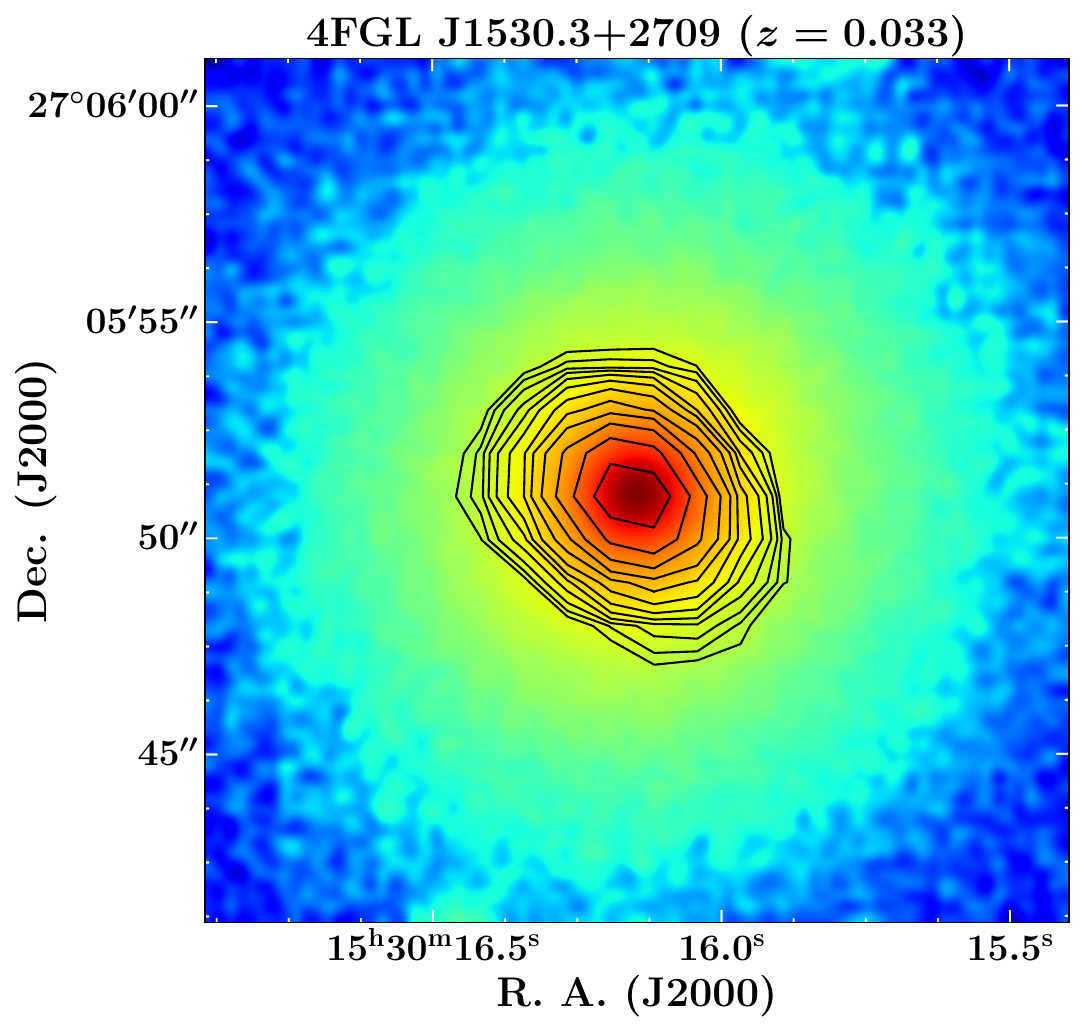}
    \includegraphics[scale=0.23]{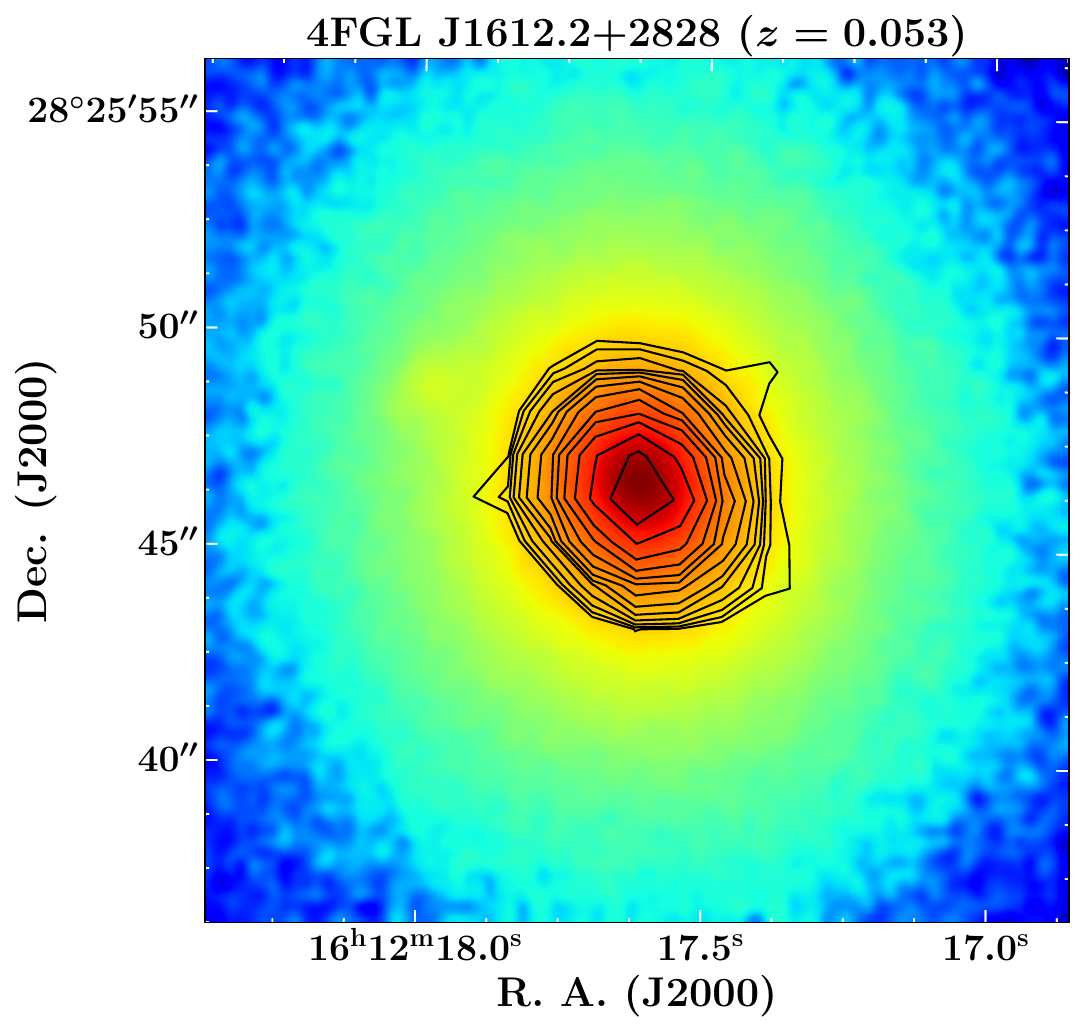}
    \includegraphics[scale=0.23]{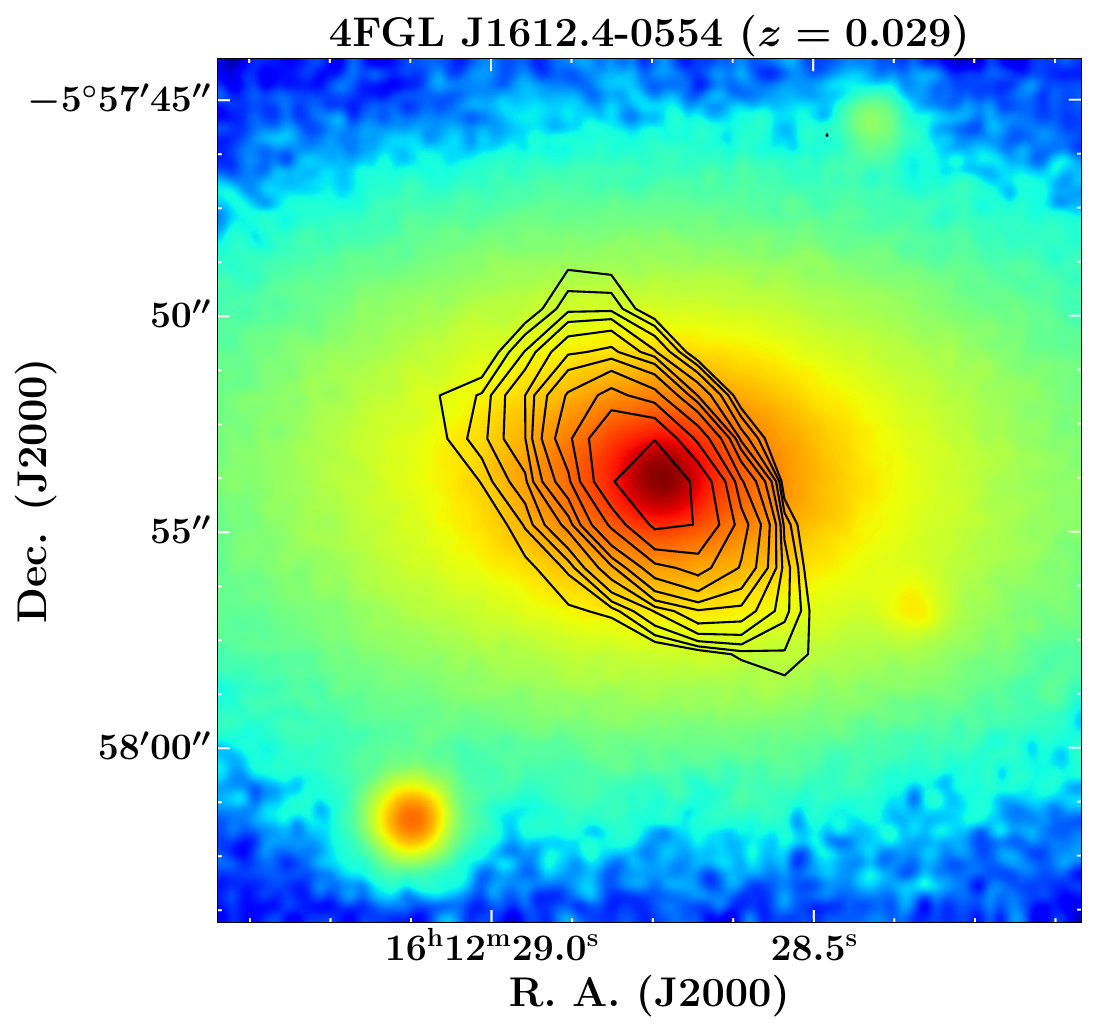}
    }
    \hbox{\hspace{2.5cm}
    \includegraphics[scale=0.25]{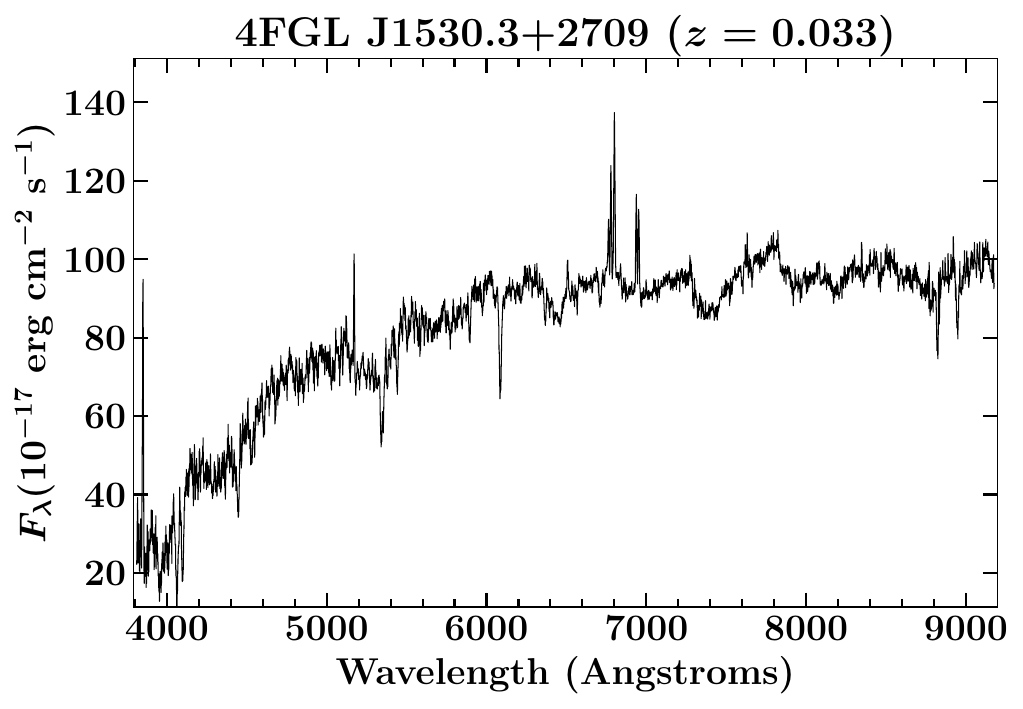}
    \includegraphics[scale=0.25]{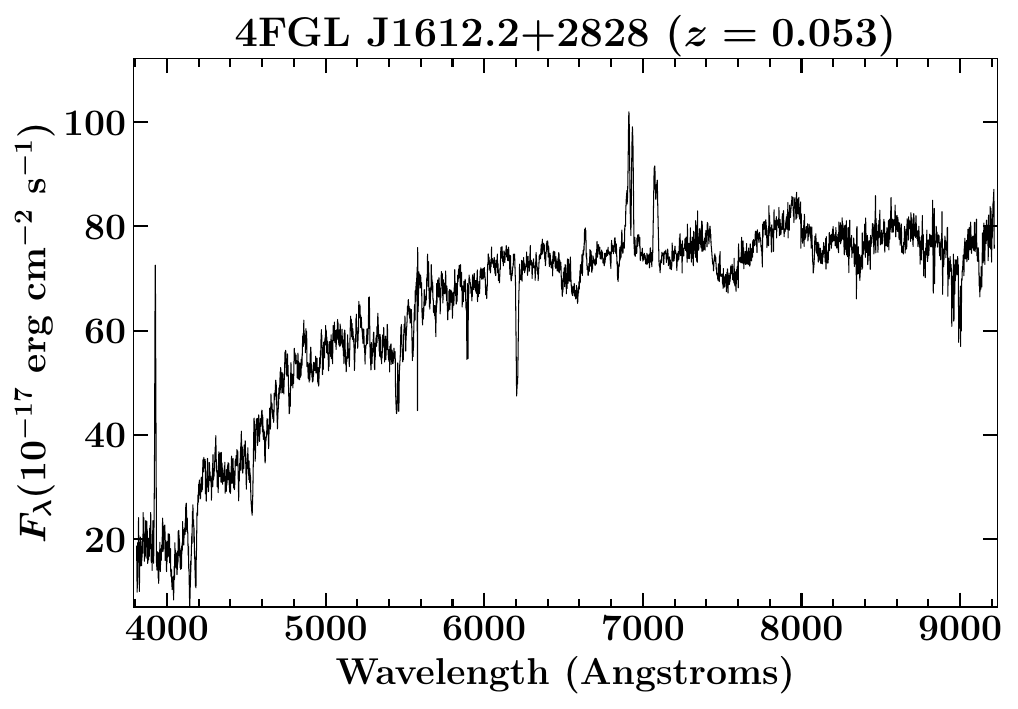}
    \includegraphics[scale=0.25]{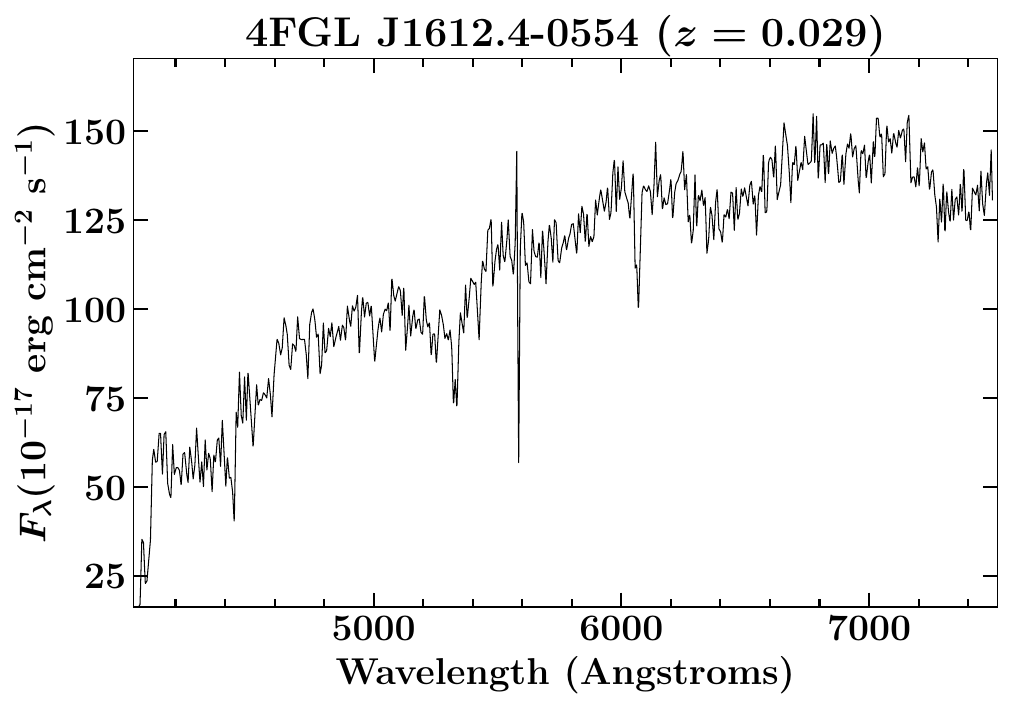}
    }
\caption{The VLASS contours overplotted on PanSTARRS $r$-band images of \gm-ray emitting FR0 radio galaxies. The contour levels start at 3$\times$rms$\times$($-\sqrt{2}$, $-1$, 1) and increases in multiple of $\sqrt{2}$. The optical spectrum of each source is also shown.}\label{fig:vlass1}
\end{figure*}

A compact radio morphology and a LEG-type spectrum are the defining properties of FR0 radio galaxies. However, these features have also been observed from a few BL Lac objects whose optical spectra are found to be dominated by the host galaxy emission \citep[e.g.,][]{1996MNRAS.281..425M,2002MNRAS.336..945L,2010AJ....139..390P}.  Therefore, for these sources, we estimated the \Ca~break strength or contrast defined as follows \citep[][]{2002MNRAS.336..945L}
\begin{equation}
\label{eq:hkbreak}
C =  \nonumber  0.14 + 0.86\left( \frac{\left< f_{\lambda,r}\right> - \left<f_{\lambda,b}\right>}{\left<f_{\lambda,r}\right>}\right)\!
\end{equation}
where $f_{\lambda,r}$ and $f_{\lambda,b}$ are the flux densities on the red (4000$-$4100 \AA) and blue (3850$-$3950 \AA) sides of the \Ca~H/K break. Furthermore, the discontinuity index, Dn(4000), can be parameterized as Dn(4000)=1/(1$-$C) \citep[][]{1999ApJ...527...54B,2002MNRAS.336..945L}. Usually, Dn(4000) and $C$ parameters have lower values for BL Lac objects, i.e., Dn(4000)$<$1.5 and $C<$0.4 \citep[][]{1987AJ.....94..899D,1991ApJS...76..813S,2010AJ....139..390P}.  Considering our sample, we then selected only those 20 objects that have Dn(4000)$>$1.7 and $C>$0.4.  

Next,  we compared their emission line properties with the {\it FR0CAT} sources on the Baldwin-Phillips-Terlevich (BPT) diagram \citep[][]{1981PASP...93....5B}. For this purpose, we employed the publicly available software `Bayesian AGN Decomposition Analysis for SDSS Spectra' \citep[{\tt BADASS},][]{2021MNRAS.500.2871S}. This tool simultaneously fits emission line and continuum features on the data and derives robust uncertainties by applying the Markov Chain Monte Carlo technique. In our work, we reproduced the continuum with the combination of power-law and empirical stellar templates to measure the stellar velocity dispersion using penalized Pixel Fitting software \citep[{\tt pPXF},][]{2017MNRAS.466..798C}. All emission lines were modeled with a Gaussian function.  We were able to measure the emission line fluxes for 9 out of 20 objects. For the remaining sources, either both H$_{\alpha}$ and H$_{\beta}$ regions were not covered in their respective spectra or the emission lines were not detected.  The resultant BPT diagnostic plot is shown in Figure~\ref{fig:bpt}. This exercise led to the rejection of three objects that are clearly separated out from the area populated by the known FR0 radio galaxies.

Finally, we searched in the literature to study the multiwavelength properties of the remaining 17 \gm-ray sources and rejected 10 objects whose broadband properties, e.g., \gm-ray variability, were found to be consistent with that exhibited by blazars. The remaining 7 AGN that have passed all the criteria adopted in this work form our final sample of \gm-ray detected FR0 radio galaxies.
 Their \gm-ray and other measured parameters are provided in Table~\ref{tab:basic_info}. One of the identified sources, 4FGL J1530.3+2709, associated with LEDA 55267 ($z=0.033$), has already been reported as a \gm-ray emitting FR0 radio galaxy by \citet[][]{2021ApJ...918L..39P} and is also present in {\it FR0CAT}. This confirms the robustness of the adopted steps to identify potential \gm-ray emitting FR0 radio galaxies.

In Figure~\ref{fig:vlass1}, we show the VLASS contours overplotted on the Panoramic Survey Telescope and Rapid Response System \citep[PanSTARRS,][]{2016arXiv161205560C} $r$-band images centered on the FR0 radio galaxies identified in this work. The optical spectra of these objects are also shown in the same plot.

\section{Discussion}\label{sec4}

We investigated the radio and optical properties of the known \gm-ray emitting AGN and identified 7 potential FR0 radio galaxies. They all exhibit compact radio morphology, and their optical spectra resemble low-excitation galaxies. Their host galaxies, as revealed by PanSTARRS $r$-band imaging, are red, early-type galaxies similar to what is typically seen in FR0 and FR I radio systems (Figure~\ref{fig:vlass1}). 

We looked into the FIRST (1.4 GHz, resolution 5 arcsec), LOFAR (144 MHz, resolution 6 arcsec), and Rapid ASKAP Continuum Survey (RACS, 887.5 MHz, resolution $\sim$15 arcsec) databases to determine the radio morphology of the identified FR0 sources at different frequencies. Both GHz and MHz frequency observations revealed the paucity of extended radio structures. These results are similar to that found by \citet[][]{2020A&A...642A.107C} who, using LOFAR observations, found $>$80\% of FR0 radio galaxies to exhibit point-like radio structures. Furthermore, we calculated the spectral index ($\alpha$, $f_\nu\propto\nu^{\alpha}$) using flux densities measured by LOFAR and/or RACS and VLASS observations and found it to be $-0.5<\alpha<0.19$ with one object, 4FGL J1530.3+2709, having an inverted radio spectrum. For a consistency check, we also inspected the NRAO VLA Sky Survey (NVSS, 1.4 GHz, resolution 45 arcsec) images of all FR0 sources with the motivation to identify any large-scale jetted emission, if any, which could be resolved out in VLASS observations due to smaller beam size. We found all 7 objects to be showing a compact morphology. These results support the FR0 nature of the reported \gm-ray emitters. 

\begin{figure}
\vbox{
    \includegraphics[width=\linewidth]{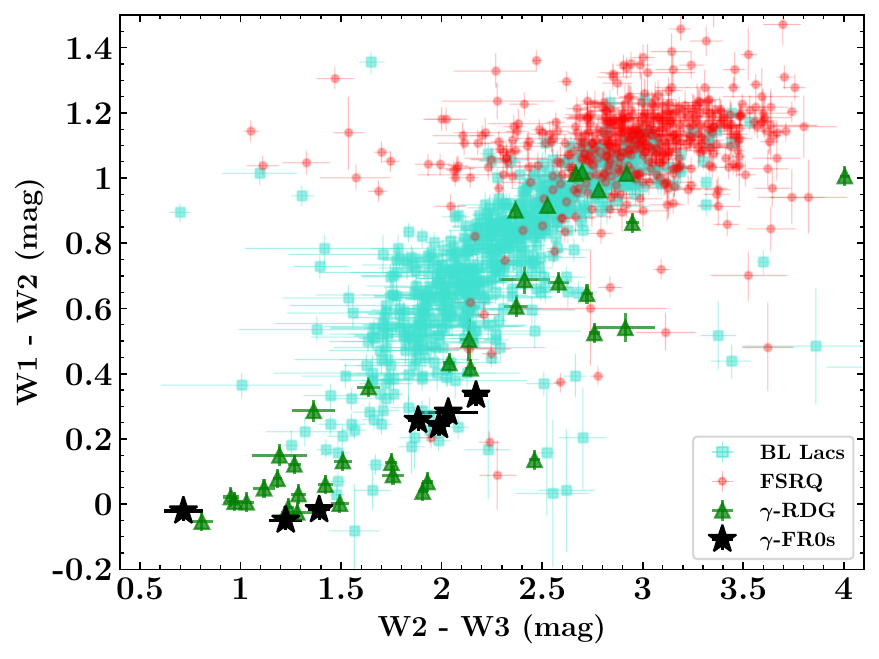}
    \includegraphics[width=\linewidth]{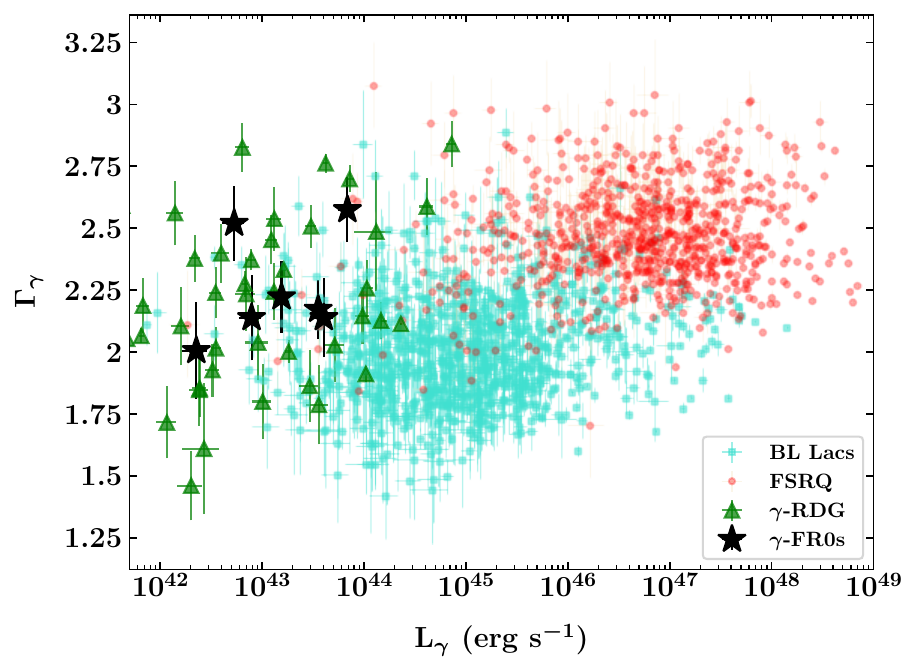}
    }
\caption{Top: The WISE colour-colour diagram showing \gm-ray detected blazars, radio galaxies, and FR0 sources as labeled. Bottom: The \gm-ray luminosity versus photon index plane for \fermi-LAT detected AGN.} \label{fig:wise}
\end{figure}

To understand the nature of FR0s, several authors have studied the milliarcsec-scale radio structures from Very Long Baseline Interferometric (VLBI) observations. For example, VLBI observations of a sample of 14 FR0s showed 6 of these to have a prominent core and two-sided jet-like structure, 5 to have a core-jet structure, and 3 to be compact \citep[][]{2018ApJ...863..155C,2021MNRAS.506.1609C}. Similarly, for 16 FR0s, \citet[][]{2023A&A...672A.104G} found 9 to have a two-sided structure. Combining eMERLIN and JVLA data, \cite{2021Galax...9..106B} found 4 of the 5 FR0s to exhibit a twin-jet structure. Also, from their European VLBI Network observations of 10 FR0s, seven appear elongated, with three of these being classified as double or triple sources and another one also being possibly two-sided. 

The VLBI-scale structures of our 7 \gm-ray emitting FR0s are available in the database maintained by Leonid Petrov\footnote{http://astrogeo.org/}. A flat-spectrum core component dominates all the sources. 4FGL J0312.4$-$3221 shows a weak component towards the east in the C-band map, while the X-band map shows a component towards the west. Further observations are required to confirm its structure. The remaining images are either compact single components or suggest a core-jet structure. There appears to be a dearth of well-defined two-sided structures in the \gm-ray selected FR0s, suggesting that these may be at smaller angles to the line of sight compared with the general population of FR0s.

In the top panel of Figure~\ref{fig:wise}, we highlight the position of \gm-ray emitting FR0 radio galaxies in the Wide-field Infrared Survey Explorer (WISE) color-color diagram where we also plot the \gm-ray detected blazars and radio galaxies \citep[e.g.,][]{2012ApJ...748...68D}. As can be seen, the FR0 objects occupy a region mainly populated by radio galaxies. Some of the FR0s also exhibit WISE colors similar to BL Lac objects, indicating contamination from the jet in the NIR band. Given that our search for FR0 radio sources began with the \gm-ray detected, hence jet-dominated sources, such overlap can be understood.

The absolute $r$-band magnitude of \gm-ray detected FR0 radio galaxies lies in the range $-21<M_{\rm r}<-23$ (Table~\ref{tab:basic_info}) similar to other FR0 sources \citep[][]{2018A&A...609A...1B}. The bolometric luminosity of the \gm-ray detected FR0 sources can be computed from \OIIIb~line luminosity following $L_{\rm bol}$=3500$L_{\rm [OIII]5007}$ \citep[][]{2004ApJ...613..109H}.  We derived the stellar velocity dispersion ($\sigma_*$) by fitting the optical spectrum of all sources with {\tt BADASS}, which uses {\tt pPXF} software to measure $\sigma_*$. The mass of the central black hole ($M_{\rm BH}$) was then estimated from $\sigma_*$ following the prescriptions of \citet[][]{2009ApJ...698..198G}. All sources have $M_{\rm BH}$ values lying in the range $8.0<\log M_{\rm BH}<9.5$ \msun~(Table~\ref{tab:basic_info}) similar to that obtained for {\it FR0CAT} objects. The accretion rate in Eddington units can be derived using the computed $M_{\rm BH}$ values and found to be $<10^{-2}$, which is similar to LEGs and other FR0 radio galaxies \citep[e.g.,][]{2012MNRAS.421.1569B,2018A&A...609A...1B}. Moreover, the discontinuity index Dn(4000) calculated for \gm-ray detected FR0s is also comparable to other FR0 and FR I radio galaxies \citep[][]{2018A&A...609A...1B}. \citet[][]{2002MNRAS.336..945L} proposed a correlation between the jet viewing angle and \Ca~break strength for a range of bulk Lorentz factor values of radio galaxies. For FR0 sources, the bulk Lorentz factor of their jets is estimated to be lower than FR Is \citep[][]{2018ApJ...863..155C,2023A&A...672A.104G}. Since all sources present in our sample have $C>0.4$, the corresponding jet viewing angle is expected to be $>$30$^{\circ}$ \citep[][]{2002MNRAS.336..945L}. 

The high-energy \gm-ray properties of FR0 radio galaxies indicate them to be low-luminosity AGN. All FR0 sources exhibit a soft \gm-ray spectrum (photon index $\Gamma_{\gamma}>$2, Figure~\ref{fig:wise}, bottom panel), which is similar to that found for the under-the-threshold population of FR0 radio galaxies employing the stacking analysis \citep[][]{2021ApJ...918L..39P}. As seen in the bottom panel of Figure~\ref{fig:wise}, FR0 objects are located in a region shared by \gm-ray detected radio galaxies. These findings suggest possibly a common \gm-ray production mechanism and that FR0s also host misaligned jets similar to FR I and II radio galaxies. Additionally, we also explored the \gm-ray variability behavior of FR0 objects using the \fermi-LAT light curve repository \citep[][]{2023ApJS..265...31A}. This repository contains the \gm-ray light curves of all sources flagged as variable in the 4FGL Catalog. None of our sources are in this repository, indicating them as non-variable \gm-ray sources. The non-variable nature could be intrinsic and could also be due to the faintness of the sources.
\section{Summary}\label{sec5}
In this work, we have identified 7 FR0 radio galaxies among the known \gm-ray emitting AGN using the data provided by ongoing multi-frequency, wide-field sky surveys, thus $\sim$quadrupling the sample size of \gm-ray detected FR0s. The broadband observational properties of these sources are found to be similar to other \gm-ray undetected FR0 radio sources \citep[][]{2018A&A...609A...1B}. Given the small sample size and being a relatively new member of the AGN family, there are several outstanding questions about their origin and nature that remain open and require dedicated studies. Future high-resolution, multi-frequency observation of this enigmatic class of radio galaxies will provide tantalizing clues about the origin of relativistic jets and their interaction with the ambient environment.

\acknowledgements
We thank the journal referee for constructive criticism that helped us improve the paper. The use of the SDSS, PanSTARRS, WISE, LOFAR, and RACS data is acknowledged. The National Radio Astronomy Observatory is a facility of the National Science Foundation operated under cooperative agreement by Associated Universities, Inc. CIRADA is funded by a grant from the Canada Foundation for Innovation 2017 Innovation Fund (Project 35999), as well as by the Provinces of Ontario, British Columbia, Alberta, Manitoba and Quebec.

\bibliographystyle{aasjournal}

\begin{thebibliography}{}
\expandafter\ifx\csname natexlab\endcsname\relax\def\natexlab#1{#1}\fi

\bibitem[{{Abdollahi} {et~al.}(2023){Abdollahi}, {Ajello}, {Baldini}, {Ballet},
  {Bastieri}, {Becerra Gonzalez}, {Bellazzini}, {Berretta}, {Bissaldi},
  {Bonino}, {Brill}, {Bruel}, {Burns}, {Buson}, {Cameron}, {Caputo}, {Caraveo},
  {Cibrario}, {Ciprini}, {Cristarella Orestano}, {Crnogorcevic}, {Cutini},
  {D'Ammando}, {De Gaetano}, {Digel}, {Di Lalla}, {Di Venere},
  {Dom{\'\i}nguez}, {Ramazani}, {Fegan}, {Ferrara}, {Fiori}, {Fleischhack},
  {Franckowiak}, {Fukazawa}, {Fusco}, {Gammaldi}, {Gargano}, {Garrappa},
  {Gasbarra}, {Gasparrini}, {Giglietto}, {Giordano}, {Giroletti}, {Green},
  {Grenier}, {Guiriec}, {Gustafsson}, {Hays}, {Horan}, {Hou},
  {J{\'o}hannesson}, {Kerr}, {Kocevski}, {Kuss}, {Latronico}, {Li}, {Liodakis},
  {Longo}, {Loparco}, {Lorusso}, {Lott}, {Lovellette}, {Lubrano}, {Maldera},
  {Manfreda}, {Mart{\'\i}-Devesa}, {Mazziotta}, {Mereu}, {Meyer}, {Michelson},
  {Mizuno}, {Monzani}, {Morselli}, {Moskalenko}, {Negro}, {Omodei}, {Orlando},
  {Ormes}, {Paneque}, {Panzarini}, {Perkins}, {Persic}, {Pesce-Rollins},
  {Pillera}, {Porter}, {Principe}, {Racusin}, {Rain{\`o}}, {Rando}, {Rani},
  {Razzano}, {Razzaque}, {Reimer}, {Reimer}, {S{\'a}nchez-Conde}, {Parkinson},
  {Scargle}, {Scotton}, {Serini}, {Sgr{\`o}}, {Siskind}, {Spandre}, {Spinelli},
  {Suson}, {Tajima}, {Thompson}, {Torres}, {Valverde}, {Venters}, {Wadiasingh},
  {Wagner}, \& {Wood}}]{2023ApJS..265...31A}
{Abdollahi}, S., {Ajello}, M., {Baldini}, L., {et~al.} 2023, \apjs, 265, 31

\bibitem[{{Abdurro'uf} {et~al.}(2022){Abdurro'uf}, {Accetta}, {Aerts}, {Silva
  Aguirre}, {Ahumada}, {Ajgaonkar}, {Filiz Ak}, {Alam}, {Allende Prieto},
  {Almeida}, {Anders}, {Anderson}, {Andrews}, {Anguiano}, {Aquino-Ort{\'\i}z},
  {Arag{\'o}n-Salamanca}, {Argudo-Fern{\'a}ndez}, {Ata}, {Aubert},
  {Avila-Reese}, {Badenes}, {Barb{\'a}}, {Barger}, {Barrera-Ballesteros},
  {Beaton}, {Beers}, {Belfiore}, {Bender}, {Bernardi}, {Bershady}, {Beutler},
  {Bidin}, {Bird}, {Bizyaev}, {Blanc}, {Blanton}, {Boardman}, {Bolton},
  {Boquien}, {Borissova}, {Bovy}, {Brandt}, {Brown}, {Brownstein}, {Brusa},
  {Buchner}, {Bundy}, {Burchett}, {Bureau}, {Burgasser}, {Cabang}, {Campbell},
  {Cappellari}, {Carlberg}, {Wanderley}, {Carrera}, {Cash}, {Chen}, {Chen},
  {Cherinka}, {Chiappini}, {Choi}, {Chojnowski}, {Chung}, {Clerc}, {Cohen},
  {Comerford}, {Comparat}, {da Costa}, {Covey}, {Crane}, {Cruz-Gonzalez},
  {Culhane}, {Cunha}, {Dai}, {Damke}, {Darling}, {Davidson}, {Davies},
  {Dawson}, {De Lee}, {Diamond-Stanic}, {Cano-D{\'\i}az}, {S{\'a}nchez},
  {Donor}, {Duckworth}, {Dwelly}, {Eisenstein}, {Elsworth}, {Emsellem},
  {Eracleous}, {Escoffier}, {Fan}, {Farr}, {Feng}, {Fern{\'a}ndez-Trincado},
  {Feuillet}, {Filipp}, {Fillingham}, {Frinchaboy}, {Fromenteau}, {Galbany},
  {Garc{\'\i}a}, {Garc{\'\i}a-Hern{\'a}ndez}, {Ge}, {Geisler}, {Gelfand},
  {G{\'e}ron}, {Gibson}, {Goddy}, {Godoy-Rivera}, {Grabowski}, {Green},
  {Greener}, {Grier}, {Griffith}, {Guo}, {Guy}, {Hadjara}, {Harding},
  {Hasselquist}, {Hayes}, {Hearty}, {Hern{\'a}ndez}, {Hill}, {Hogg},
  {Holtzman}, {Horta}, {Hsieh}, {Hsu}, {Hsu}, {Huber}, {Huertas-Company},
  {Hutchinson}, {Hwang}, {Ibarra-Medel}, {Chitham}, {Ilha}, {Imig}, {Jaekle},
  {Jayasinghe}, {Ji}, {Johnson}, {Jones}, {J{\"o}nsson}, {Katkov}, {Khalatyan},
  {Kinemuchi}, {Kisku}, {Knapen}, {Kneib}, {Kollmeier}, {Kong}, {Kounkel},
  {Kreckel}, {Krishnarao}, {Lacerna}, {Lane}, {Langgin}, {Lavender}, {Law},
  {Lazarz}, {Leung}, {Leung}, {Lewis}, {Li}, {Li}, {Lian}, {Liang}, {Lin},
  {Lin}, {Lin}, {Lintott}, {Long}, {Longa-Pe{\~n}a}, {L{\'o}pez-Cob{\'a}},
  {Lu}, {Lundgren}, {Luo}, {Mackereth}, {de la Macorra}, {Mahadevan},
  {Majewski}, {Manchado}, {Mandeville}, {Maraston}, {Margalef-Bentabol},
  {Masseron}, {Masters}, {Mathur}, {McDermid}, {Mckay}, {Merloni},
  {Merrifield}, {Meszaros}, {Miglio}, {Di Mille}, {Minniti}, {Minsley},
  {Monachesi}, {Moon}, {Mosser}, {Mulchaey}, {Muna}, {Mu{\~n}oz}, {Myers},
  {Myers}, {Nadathur}, {Nair}, {Nandra}, {Neumann}, {Newman}, {Nidever},
  {Nikakhtar}, {Nitschelm}, {O'Connell}, {Garma-Oehmichen}, {Luan Souza de
  Oliveira}, {Olney}, {Oravetz}, {Ortigoza-Urdaneta}, {Osorio}, {Otter},
  {Pace}, {Padilla}, {Pan}, {Pan}, {Parikh}, {Parker}, {Peirani}, {Pe{\~n}a
  Ram{\'\i}rez}, {Penny}, {Percival}, {Perez-Fournon}, {Pinsonneault},
  {Poidevin}, {Poovelil}, {Price-Whelan}, {B{\'a}rbara de Andrade Queiroz},
  {Raddick}, {Ray}, {Rembold}, {Riddle}, {Riffel}, {Riffel}, {Rix}, {Robin},
  {Rodr{\'\i}guez-Puebla}, {Roman-Lopes}, {Rom{\'a}n-Z{\'u}{\~n}iga}, {Rose},
  {Ross}, {Rossi}, {Rubin}, {Salvato}, {S{\'a}nchez}, {S{\'a}nchez-Gallego},
  {Sanderson}, {Santana Rojas}, {Sarceno}, {Sarmiento}, {Sayres}, {Sazonova},
  {Schaefer}, {Schiavon}, {Schlegel}, {Schneider}, {Schultheis}, {Schwope},
  {Serenelli}, {Serna}, {Shao}, {Shapiro}, {Sharma}, {Shen}, {Shetrone}, {Shu},
  {Simon}, {Skrutskie}, {Smethurst}, {Smith}, {Sobeck}, {Spoo}, {Sprague},
  {Stark}, {Stassun}, {Steinmetz}, {Stello}, {Stone-Martinez},
  {Storchi-Bergmann}, {Stringfellow}, {Stutz}, {Su}, {Taghizadeh-Popp},
  {Talbot}, {Tayar}, {Telles}, {Teske}, {Thakar}, {Theissen}, {Tkachenko},
  {Thomas}, {Tojeiro}, {Hernandez Toledo}, {Troup}, {Trump}, {Trussler},
  {Turner}, {Tuttle}, {Unda-Sanzana}, {V{\'a}zquez-Mata}, {Valentini},
  {Valenzuela}, {Vargas-Gonz{\'a}lez}, {Vargas-Maga{\~n}a}, {Alfaro},
  {Villanova}, {Vincenzo}, {Wake}, {Warfield}, {Washington}, {Weaver},
  {Weijmans}, {Weinberg}, {Weiss}, {Westfall}, {Wild}, {Wilde}, {Wilson},
  {Wilson}, {Wilson}, {Wolf}, {Wood-Vasey}, {Yan}, {Zamora}, {Zasowski},
  {Zhang}, {Zhao}, {Zheng}, {Zheng}, \& {Zhu}}]{2022ApJS..259...35A}
{Abdurro'uf}, {Accetta}, K., {Aerts}, C., {et~al.} 2022, \apjs, 259, 35

\bibitem[{{Ajello} {et~al.}(2022){Ajello}, {Baldini}, {Ballet}, {Bastieri},
  {Becerra Gonzalez}, {Bellazzini}, {Berretta}, {Bissaldi}, {Bonino}, {Brill},
  {Bruel}, {Buson}, {Caputo}, {Caraveo}, {Cheung}, {Chiaro}, {Cibrario},
  {Ciprini}, {Crnogorcevic}, {Cutini}, {D'Ammando}, {De Gaetano}, {Di Lalla},
  {Di Venere}, {Dom{\'\i}nguez}, {Ramazani}, {Ferrara}, {Fiori}, {Fukazawa},
  {Funk}, {Fusco}, {Gammaldi}, {Gargano}, {Garrappa}, {Gasparrini},
  {Giglietto}, {Giordano}, {Giroletti}, {Green}, {Grenier}, {Guiriec}, {Horan},
  {Hou}, {Kayanoki}, {Kuss}, {Larsson}, {Latronico}, {Lewis}, {Li}, {Liodakis},
  {Longo}, {Loparco}, {Lott}, {Lovellette}, {Lubrano}, {Madejski}, {Maldera},
  {Manfreda}, {Mart{\'\i}-Devesa}, {Mazziotta}, {Mereu}, {Michelson},
  {Mirabal}, {Mitthumsiri}, {Mizuno}, {Monzani}, {Morselli}, {Moskalenko},
  {Negro}, {Ojha}, {Orienti}, {Orlando}, {Ormes}, {Pei}, {Pe{\~n}a-Herazo},
  {Persic}, {Pesce-Rollins}, {Petrosian}, {Pillera}, {Poon}, {Porter},
  {Principe}, {Rain{\`o}}, {Rando}, {Rani}, {Razzano}, {Razzaque}, {Reimer},
  {Reimer}, {Scotton}, {Serini}, {Sgr{\`o}}, {Siskind}, {Spandre}, {Spinelli},
  {Suson}, {Tajima}, {Torres}, {Valverde}, {Yassin}, \&
  {Zaharijas}}]{2022ApJS..263...24A}
{Ajello}, M., {Baldini}, L., {Ballet}, J., {et~al.} 2022, \apjs, 263, 24

\bibitem[{{Baldi}(2023)}]{2023A&ARv..31....3B}
{Baldi}, R.~D. 2023, \aapr, 31, 3

\bibitem[{{Baldi} \& {Capetti}(2009)}]{2009A&A...508..603B}
{Baldi}, R.~D., \& {Capetti}, A. 2009, \aap, 508, 603

\bibitem[{{Baldi} \& {Capetti}(2010)}]{2010A&A...519A..48B}
---. 2010, \aap, 519, A48

\bibitem[{{Baldi} {et~al.}(2019){Baldi}, {Capetti}, \&
  {Giovannini}}]{2019MNRAS.482.2294B}
{Baldi}, R.~D., {Capetti}, A., \& {Giovannini}, G. 2019, \mnras, 482, 2294

\bibitem[{{Baldi} {et~al.}(2018){Baldi}, {Capetti}, \&
  {Massaro}}]{2018A&A...609A...1B}
{Baldi}, R.~D., {Capetti}, A., \& {Massaro}, F. 2018, \aap, 609, A1

\bibitem[{{Baldi} {et~al.}(2021){Baldi}, {Giovannini}, \&
  {Capetti}}]{2021Galax...9..106B}
{Baldi}, R.~D., {Giovannini}, G., \& {Capetti}, A. 2021, Galaxies, 9, 106

\bibitem[{{Baldwin} {et~al.}(1981){Baldwin}, {Phillips}, \&
  {Terlevich}}]{1981PASP...93....5B}
{Baldwin}, J.~A., {Phillips}, M.~M., \& {Terlevich}, R. 1981, \pasp, 93, 5

\bibitem[{{Balogh} {et~al.}(1999){Balogh}, {Morris}, {Yee}, {Carlberg}, \&
  {Ellingson}}]{1999ApJ...527...54B}
{Balogh}, M.~L., {Morris}, S.~L., {Yee}, H.~K.~C., {Carlberg}, R.~G., \&
  {Ellingson}, E. 1999, \apj, 527, 54

\bibitem[{{Best} \& {Heckman}(2012)}]{2012MNRAS.421.1569B}
{Best}, P.~N., \& {Heckman}, T.~M. 2012, \mnras, 421, 1569

\bibitem[{{Brinchmann} {et~al.}(2004){Brinchmann}, {Charlot}, {White},
  {Tremonti}, {Kauffmann}, {Heckman}, \& {Brinkmann}}]{2004MNRAS.351.1151B}
{Brinchmann}, J., {Charlot}, S., {White}, S.~D.~M., {et~al.} 2004, \mnras, 351,
  1151

\bibitem[{{Capetti} {et~al.}(2020){Capetti}, {Brienza}, {Baldi}, {Giovannini},
  {Morganti}, {Hardcastle}, {Rottgering}, {Brunetti}, {Best}, \&
  {Miley}}]{2020A&A...642A.107C}
{Capetti}, A., {Brienza}, M., {Baldi}, R.~D., {et~al.} 2020, \aap, 642, A107

\bibitem[{{Cappellari}(2017)}]{2017MNRAS.466..798C}
{Cappellari}, M. 2017, \mnras, 466, 798

\bibitem[{{Chambers} {et~al.}(2016){Chambers}, {Magnier}, {Metcalfe},
  {Flewelling}, {Huber}, {Waters}, {Denneau}, {Draper}, {Farrow}, {Finkbeiner},
  {Holmberg}, {Koppenhoefer}, {Price}, {Rest}, {Saglia}, {Schlafly}, {Smartt},
  {Sweeney}, {Wainscoat}, {Burgett}, {Chastel}, {Grav}, {Heasley}, {Hodapp},
  {Jedicke}, {Kaiser}, {Kudritzki}, {Luppino}, {Lupton}, {Monet}, {Morgan},
  {Onaka}, {Shiao}, {Stubbs}, {Tonry}, {White}, {Ba{\~n}ados}, {Bell},
  {Bender}, {Bernard}, {Boegner}, {Boffi}, {Botticella}, {Calamida},
  {Casertano}, {Chen}, {Chen}, {Cole}, {Deacon}, {Frenk}, {Fitzsimmons},
  {Gezari}, {Gibbs}, {Goessl}, {Goggia}, {Gourgue}, {Goldman}, {Grant},
  {Grebel}, {Hambly}, {Hasinger}, {Heavens}, {Heckman}, {Henderson}, {Henning},
  {Holman}, {Hopp}, {Ip}, {Isani}, {Jackson}, {Keyes}, {Koekemoer}, {Kotak},
  {Le}, {Liska}, {Long}, {Lucey}, {Liu}, {Martin}, {Masci}, {McLean}, {Mindel},
  {Misra}, {Morganson}, {Murphy}, {Obaika}, {Narayan}, {Nieto-Santisteban},
  {Norberg}, {Peacock}, {Pier}, {Postman}, {Primak}, {Rae}, {Rai}, {Riess},
  {Riffeser}, {Rix}, {R{\"o}ser}, {Russel}, {Rutz}, {Schilbach}, {Schultz},
  {Scolnic}, {Strolger}, {Szalay}, {Seitz}, {Small}, {Smith}, {Soderblom},
  {Taylor}, {Thomson}, {Taylor}, {Thakar}, {Thiel}, {Thilker}, {Unger},
  {Urata}, {Valenti}, {Wagner}, {Walder}, {Walter}, {Watters}, {Werner},
  {Wood-Vasey}, \& {Wyse}}]{2016arXiv161205560C}
{Chambers}, K.~C., {Magnier}, E.~A., {Metcalfe}, N., {et~al.} 2016, arXiv
  e-prints, arXiv:1612.05560

\bibitem[{{Cheng} {et~al.}(2021){Cheng}, {An}, {Sohn}, {Hong}, \&
  {Wang}}]{2021MNRAS.506.1609C}
{Cheng}, X., {An}, T., {Sohn}, B.~W., {Hong}, X., \& {Wang}, A. 2021, \mnras,
  506, 1609

\bibitem[{{Cheng} \& {An}(2018)}]{2018ApJ...863..155C}
{Cheng}, X.~P., \& {An}, T. 2018, \apj, 863, 155

\bibitem[{{D'Abrusco} {et~al.}(2012){D'Abrusco}, {Massaro}, {Ajello},
  {Grindlay}, {Smith}, \& {Tosti}}]{2012ApJ...748...68D}
{D'Abrusco}, R., {Massaro}, F., {Ajello}, M., {et~al.} 2012, \apj, 748, 68

\bibitem[{{Dressler} \& {Shectman}(1987)}]{1987AJ.....94..899D}
{Dressler}, A., \& {Shectman}, S.~A. 1987, \aj, 94, 899

\bibitem[{{Fanaroff} \& {Riley}(1974)}]{1974MNRAS.167P..31F}
{Fanaroff}, B.~L., \& {Riley}, J.~M. 1974, \mnras, 167, 31P

\bibitem[{{Ghisellini}(2011)}]{2011AIPC.1381..180G}
{Ghisellini}, G. 2011, in American Institute of Physics Conference Series, Vol.
  1381, 25th Texas Symposium on Relativistic AstroPhysics (Texas 2010), ed.
  F.~A. {Aharonian}, W.~{Hofmann}, \& F.~M. {Rieger}, 180--198

\bibitem[{{Giovannini} {et~al.}(2023){Giovannini}, {Baldi}, {Capetti},
  {Giroletti}, \& {Lico}}]{2023A&A...672A.104G}
{Giovannini}, G., {Baldi}, R.~D., {Capetti}, A., {Giroletti}, M., \& {Lico}, R.
  2023, \aap, 672, A104

\bibitem[{{Grandi} {et~al.}(2016){Grandi}, {Capetti}, \&
  {Baldi}}]{2016MNRAS.457....2G}
{Grandi}, P., {Capetti}, A., \& {Baldi}, R.~D. 2016, \mnras, 457, 2

\bibitem[{{G{\"u}ltekin} {et~al.}(2009){G{\"u}ltekin}, {Richstone}, {Gebhardt},
  {Lauer}, {Tremaine}, {Aller}, {Bender}, {Dressler}, {Faber}, {Filippenko},
  {Green}, {Ho}, {Kormendy}, {Magorrian}, {Pinkney}, \&
  {Siopis}}]{2009ApJ...698..198G}
{G{\"u}ltekin}, K., {Richstone}, D.~O., {Gebhardt}, K., {et~al.} 2009, \apj,
  698, 198

\bibitem[{{Heckman} {et~al.}(2004){Heckman}, {Kauffmann}, {Brinchmann},
  {Charlot}, {Tremonti}, \& {White}}]{2004ApJ...613..109H}
{Heckman}, T.~M., {Kauffmann}, G., {Brinchmann}, J., {et~al.} 2004, \apj, 613,
  109

\bibitem[{{Helfand} {et~al.}(2015){Helfand}, {White}, \&
  {Becker}}]{2015ApJ...801...26H}
{Helfand}, D.~J., {White}, R.~L., \& {Becker}, R.~H. 2015, \apj, 801, 26

\bibitem[{{Kewley} {et~al.}(2006){Kewley}, {Groves}, {Kauffmann}, \&
  {Heckman}}]{2006MNRAS.372..961K}
{Kewley}, L.~J., {Groves}, B., {Kauffmann}, G., \& {Heckman}, T. 2006, \mnras,
  372, 961

\bibitem[{{Lacy} {et~al.}(2020){Lacy}, {Baum}, {Chandler}, {Chatterjee},
  {Clarke}, {Deustua}, {English}, {Farnes}, {Gaensler}, {Gugliucci},
  {Hallinan}, {Kent}, {Kimball}, {Law}, {Lazio}, {Marvil}, {Mao}, {Medlin},
  {Mooley}, {Murphy}, {Myers}, {Osten}, {Richards}, {Rosolowsky}, {Rudnick},
  {Schinzel}, {Sivakoff}, {Sjouwerman}, {Taylor}, {White}, {Wrobel},
  {Andernach}, {Beasley}, {Berger}, {Bhatnager}, {Birkinshaw}, {Bower},
  {Brandt}, {Brown}, {Burke-Spolaor}, {Butler}, {Comerford}, {Demorest}, {Fu},
  {Giacintucci}, {Golap}, {G{\"u}th}, {Hales}, {Hiriart}, {Hodge}, {Horesh},
  {Ivezi{\'c}}, {Jarvis}, {Kamble}, {Kassim}, {Liu}, {Loinard}, {Lyons},
  {Masters}, {Mezcua}, {Moellenbrock}, {Mroczkowski}, {Nyland}, {O'Dea},
  {O'Sullivan}, {Peters}, {Radford}, {Rao}, {Robnett}, {Salcido}, {Shen},
  {Sobotka}, {Witz}, {Vaccari}, {van Weeren}, {Vargas}, {Williams}, \&
  {Yoon}}]{2020PASP..132c5001L}
{Lacy}, M., {Baum}, S.~A., {Chandler}, C.~J., {et~al.} 2020, \pasp, 132, 035001

\bibitem[{{Landt} {et~al.}(2002){Landt}, {Padovani}, \&
  {Giommi}}]{2002MNRAS.336..945L}
{Landt}, H., {Padovani}, P., \& {Giommi}, P. 2002, \mnras, 336, 945

\bibitem[{{Marcha} {et~al.}(1996){Marcha}, {Browne}, {Impey}, \&
  {Smith}}]{1996MNRAS.281..425M}
{Marcha}, M.~J.~M., {Browne}, I.~W.~A., {Impey}, C.~D., \& {Smith}, P.~S. 1996,
  \mnras, 281, 425

\bibitem[{{Merten} {et~al.}(2021){Merten}, {Boughelilba}, {Reimer}, {Da Vela},
  {Vorobiov}, {Tavecchio}, {Bonnoli}, {Lundquist}, \&
  {Righi}}]{2021APh...12802564M}
{Merten}, L., {Boughelilba}, M., {Reimer}, A., {et~al.} 2021, Astroparticle
  Physics, 128, 102564

\bibitem[{{Mingo} {et~al.}(2019){Mingo}, {Croston}, {Hardcastle}, {Best},
  {Duncan}, {Morganti}, {Rottgering}, {Sabater}, {Shimwell}, {Williams},
  {Brienza}, {Gurkan}, {Mahatma}, {Morabito}, {Prandoni}, {Bondi}, {Ineson}, \&
  {Mooney}}]{2019MNRAS.488.2701M}
{Mingo}, B., {Croston}, J.~H., {Hardcastle}, M.~J., {et~al.} 2019, \mnras, 488,
  2701

\bibitem[{{Paliya}(2021)}]{2021ApJ...918L..39P}
{Paliya}, V.~S. 2021, \apjl, 918, L39

\bibitem[{{Paliya} {et~al.}(2021){Paliya}, {Dom{\'\i}nguez}, {Ajello},
  {Olmo-Garc{\'\i}a}, \& {Hartmann}}]{2021ApJS..253...46P}
{Paliya}, V.~S., {Dom{\'\i}nguez}, A., {Ajello}, M., {Olmo-Garc{\'\i}a}, A., \&
  {Hartmann}, D. 2021, \apjs, 253, 46

\bibitem[{{Plotkin} {et~al.}(2010){Plotkin}, {Anderson}, {Brandt},
  {Diamond-Stanic}, {Fan}, {Hall}, {Kimball}, {Richmond}, {Schneider},
  {Shemmer}, {Voges}, {York}, {Bahcall}, {Snedden}, {Bizyaev}, {Brewington},
  {Malanushenko}, {Malanushenko}, {Oravetz}, {Pan}, \&
  {Simmons}}]{2010AJ....139..390P}
{Plotkin}, R.~M., {Anderson}, S.~F., {Brandt}, W.~N., {et~al.} 2010, \aj, 139,
  390

\bibitem[{{Rees}(1966)}]{1966Natur.211..468R}
{Rees}, M.~J. 1966, \nat, 211, 468

\bibitem[{{Sexton} {et~al.}(2021){Sexton}, {Matzko}, {Darden}, {Canalizo}, \&
  {Gorjian}}]{2021MNRAS.500.2871S}
{Sexton}, R.~O., {Matzko}, W., {Darden}, N., {Canalizo}, G., \& {Gorjian}, V.
  2021, \mnras, 500, 2871

\bibitem[{{Stocke} {et~al.}(1991){Stocke}, {Morris}, {Gioia}, {Maccacaro},
  {Schild}, {Wolter}, {Fleming}, \& {Henry}}]{1991ApJS...76..813S}
{Stocke}, J.~T., {Morris}, S.~L., {Gioia}, I.~M., {et~al.} 1991, \apjs, 76, 813

\bibitem[{{Tavecchio} {et~al.}(2018){Tavecchio}, {Righi}, {Capetti}, {Grandi},
  \& {Ghisellini}}]{2018MNRAS.475.5529T}
{Tavecchio}, F., {Righi}, C., {Capetti}, A., {Grandi}, P., \& {Ghisellini}, G.
  2018, \mnras, 475, 5529

\bibitem[{{Torresi} {et~al.}(2018){Torresi}, {Grandi}, {Capetti}, {Baldi}, \&
  {Giovannini}}]{2018MNRAS.476.5535T}
{Torresi}, E., {Grandi}, P., {Capetti}, A., {Baldi}, R.~D., \& {Giovannini}, G.
  2018, \mnras, 476, 5535

\end{thebibliography}

\end{document}